**Flux-trapping experiments on ultra-high pressure hydrides as evidence of superconductivity**

Jeffery L. Tallon

Robinson Research Institute, Victoria University of Wellington, PO Box 33436, Lower Hutt 5046, New Zealand

**Abstract:** Flux trapping magnetisation studies on $H_3S$ at ultra-high pressure have been reported by Minkov *et al*. as definitive evidence of superconductivity in this hydride system. This is very helpful in a field that has become somewhat controversial. However, this conclusion has been questioned based on an apparent zero-field cooled (ZFC) *linear* magnetisation at low field. The standard Bean model would require an approximately quadratic dependence. In support, we note that the reported ZFC magnetisation is indeed super-linear and consistent both with model calculations for thin discs and with the ZFC magnetisation reported for $YBa_2Cu_3O_y$ films. We conclude that the reported high-pressure magnetisation data is fully consistent with superconductivity and that there is no reason, in this particular data set, to reject the original inference of hydride superconductivity.

The question of high-temperature superconductivity in hydrides at ultra-high pressure has become somewhat controversial following the retraction of certain high-profile publications [1-3]. Some authors remain optimistic about the field, as indeed do I, while others argue that there is no conclusive evidence of superconductivity in high-pressure hydrides [4]. It seems fair to say that, in light of these retractions and controversy, the standard of proof needs to be raised in order to maintain, or recover, credibility.

To this end, Minkov *et al*. [5], took an important step by measuring trapped magnetic flux in field-cooled (FC) and zero-field-cooled (ZFC) samples of $H_3S$ at a pressure of 130 GPa and a temperature of 30 K. The data was consistent, they claim, with the expectations of the much-used Bean model of flux trapping in superconductors, and hence that their samples were indeed superconducting.

Have Minkov *et al*. met the bar? Some authors think not [6,7]. They rightly state that the trapped moment for low field should increase *quadratically* with field in the ZFC protocol and *linearly* with field in the FC protocol if it is indeed a signature of trapped flux resulting from supercurrents [7]. They argued that the ZFC magnetisation data of Minkov *et al*. [5] increases linearly with field and hence is inconsistent with the expected behavior of superconductors. Several other known superconductors were examined to illustrate the canonical initial linear slope under field cooling (FC) contrasting the initial quadratic behavior under ZFC. However, the ZFC magnetisation data of Minkov *et al*. for $H_3S$ is clearly superlinear, so let us see how good the fit is to the Bean model and how well it compares with other well-known superconductors such as $YBa_2Cu_3O_{7-\delta}$.

Clem and Sanchez [8] have examined the FC and ZFC magnetisation of a thin disc, and the field-dependence to saturation has been computed [6]. Fig. 1 shows the calculated trapped magnetic moment (red dashed curve in the main panel and small red crosses in the insert panel). The measured data of Minkov *et al*. for $H_3S$ is plotted by the squares and error bars. One can see that the fit is excellent, the only adjustable parameters being $H_p$, the penetration field (= 0.024 tesla), the magnitude of the saturation moment at high field, $m_s$, and the full penetration field, $H^*$. The excellence of the agreement (given the size of the error bars) can be seen in the insert panel at low field where the super-linear behaviour of the raw data is evident to the eye. Certainly, there is nothing in the comparison of model and data that could lead one to reject the interpretation of superconductivity in $H_3S$.

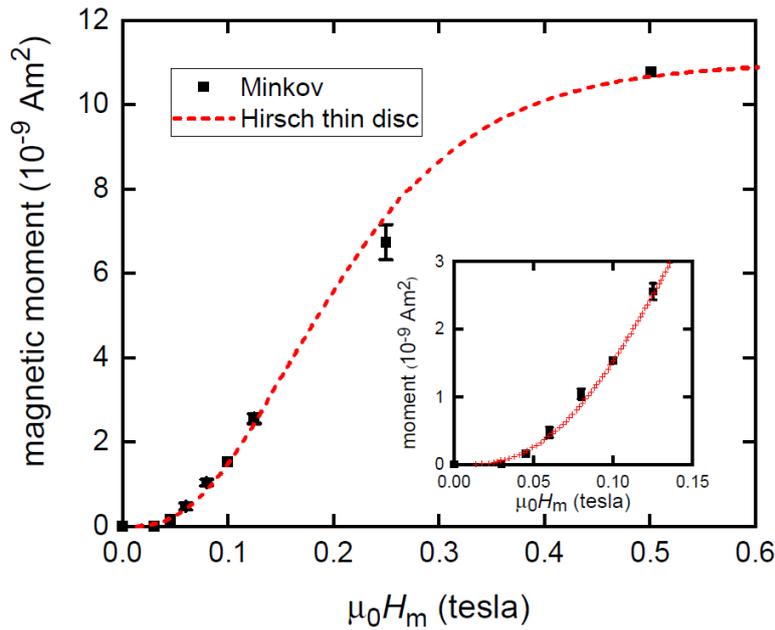

Figure 1. The ZFC trapped magnetic moment for $H_3S$ at 30 K and 155 GPa measured by Minkov et al. [5] (black square data points and error bars). The red dashed curve (and small red crosses in the low-field insert panel) is the ZFC trapped moment for a thin disc [7,8]. The comparison is made by selecting a penetration field $\mu_0 H_p$ = 0.2 tesla.

It is also helpful to compare with the reported field-dependent trapped flux in $YBa_2Cu_3O_{7-\delta}$. Müller et al. [9] investigated the magnetisation of granular samples of $Bi_2Sr_2CaCu_2O_{8+\delta}$ and compared with the measured magnetisation of coupled arrays of thin films of $YBa_2Cu_3O_{7-\delta}$ as a model granular system. Their reported behavior of individual uncoupled-array films of $YBa_2Cu_3O_{7-\delta}$ under ZFC serve as a useful comparison with the data of Minkov et al. The $YBa_2Cu_3O_{7-\delta}$ data is reproduced in Fig. 2 by the red curve in the main panel, and the small red asterisks in the insert panel showing the low-field behavior. The data of Minkov et al. is reproduced by the black squares and error bars. Again, with a penetration field of 0.024 tesla the match at low field is excellent. Across the entire field range (main panel) the fit is not so good, though it can be improved by increasing slightly the magnitude of the full penetration field, $H^*$ (see blue dashed curve). Any small discrepancies are readily attributable to a field-dependent critical current density and/or a *distributed* granularity (see Note on Granularity). If one is to dismiss superconductivity in $H_3S$ based on the ZFC magnetic flux trapping, then one would have to dismiss superconductivity in $YBa_2Cu_3O_{7-\delta}$ under the same terms. The most straightforward interpretation of the flux-trapping data of Minkov et al. is, as they claim, superconductivity. Of course, further ultra-high-pressure flux-trapping studies on other hydrides and other superconductors will help resolve the issue more definitively. In the meantime, the case for superconductivity in hydrides, based on the flux-trapping behavior of $H_3S$ as shown in Figs. 1 and 2, would appear to be rather strong.

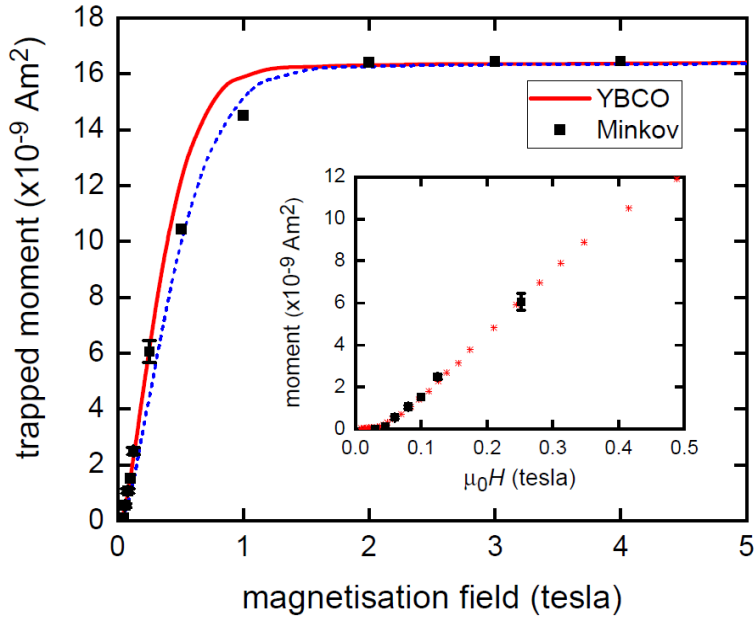

Figure 2. The ZFC trapped magnetic moment for $H_3S$ at 30 K and 155 GPa measured by Minkov *et al.* [5] (black square data points and error bars). The red dashed curve (and small red asterisks in the low-field insert panel) is the ZFC trapped moment for an isolated film of $YBa_2Cu_3O_{7-\delta}$ reported by Mueller *et al.* [9]. The blue dashed curve is a comparison where the magnitude of the full penetration field, $H^*$, has been made slightly higher than for the red curve.

**Note on Granularity**

The fits to the experimental data in Figs. 1 and 2, while excellent at lower field, show some deviations at higher field. This is to be expected if the samples are granular. A crossover from large inter-grain shielding currents at low-field to small intra-grain supercurrents at high field naturally leads to a two-step behavior of trapped flux versus field. Such two-step behavior can be seen in the measurements of Mueller *et al.* [9] on Josephson-coupled thin-film arrays, but is not apparent in the measurements of Minkov *et al.* However, such a crossover would be smeared out if the granularity is distributed. We may estimate the effects of granularity as follows. For a thin disc of radius $R$, thickness $h$ and critical current density, $J_c$, we have $m_s = \pi J_c R^3 h/2$ and $H^* = J_c h/2$ [8]. Thus, a field-dependent critical current density may not greatly alter the shape of the magnetization curve (it reduces both $m_s$ and $H^*$). On the other hand, a distributed granularity results in an *effective* radius which decreases with field and so flattens the magnetisation curve by decreasing $m_s$ alone. In such a model one must multiply the magnetisation per grain by the effective number of grains, which goes as $(R_0/R)^2$. Thus, $m_s$ will trend as $\pi J_c R_0^2 R(H) h/2$, where $R_0 = R(H=0)$. In the presence of distributed granularity, a smooth crossover of $R_{eff}$ from $R_0$ to some smaller average high-field value, $R_\infty$, clearly results in a smooth reduction in $m_s$ thus resulting in a degree of flattening of the trapped-flux curve, as observed in the data of Minkov *et al*.